\documentclass[useAMS,usenatbib,intlimits,centertags]{mn2e}
\usepackage{graphicx}
\usepackage{amsmath,amssymb}
\usepackage{booktabs}
\usepackage{dcolumn} 
\usepackage{array}
\newcommand{\PBS}[1]{\let\temp=\\#1\let\\=\temp}

\allowdisplaybreaks
\renewcommand{\d}{\text{d}}
\newcommand{\e}{\text{e}}
\newcommand{\G}{\text{G}}

\begin{document}

 \title{Uniformly Rotating Homogeneous Rings in post-Newtonian Gravity}

 \author[S.\ Horatschek \& D.\ Petroff]
 {Stefan Horatschek\footnotemark\addtocounter{footnote}{-1}
 and David Petroff\thanks{E-mail: {\tt S.Horatschek@tpi.uni-jena.de} (SH);\newline
 {\tt D.Petroff@tpi.uni-jena.de} (DP)}\\
 Theoretisch-Physikalisches Institut, University of Jena, Max-Wien-Platz 1, 07743 Jena, Germany}
 \date{\today}

 \pagerange{\pageref{firstpage}--\pageref{lastpage}} \pubyear{2010}

 \maketitle

 \label{firstpage}

 \begin{abstract}
  In this paper uniformly rotating relativistic rings are investigated analytically utilizing two different approximations simultaneously: (1) an expansion about the thin ring limit (the cross-section is small compared with the size of the whole ring) (2) post-Newtonian expansions. The analytic results for rings are compared with numerical solutions.
 \end{abstract}

\begin{keywords} gravitation -- methods: analytical -- stars: rotation. \end{keywords}

\section{Introduction}
The problem of self-gravitating, axially symmetric rings in equilibrium
can be tackled in Newtonian gravity by expanding it about the thin ring limit,
where the cross-section of the ring tends to a circle.
Doing so, \citet{Kowalewsky85}, \citet{Poincare85b,Poincare85c,Poincare85d} and \citet{Dyson92, Dyson93}
obtained series for homogeneous rings and \citet{Ostriker64,Ostriker64b,Ostriker65} and \citet{PH08b}
for polytropic rings. Alternatively by using a Roche model \citep{Roche73}
rings with a sufficiently soft equation of state can be described approximately, see \citet{HP09}.
\par

With the help of numerical methods it was possible to study homogeneous rings and their connection to the Maclaurin spheroids \citep*{Wong74,ES81,EH85,AKM03} as well as non-homogeneous rings \citep*{Hachisu86}.
Furthermore relativistic rings and their transition to the extreme Kerr Black Hole
\citep*{AKM03c,FHA05,LPA07} can be calculated to near-machine accuracy.
\par

However no analytic work has been done for relativistic rings. The reason may be that because of
the non-linearity of Einstein's field equations the expansion about the thin ring limit does not work.
In contrast, if one additionally expands the rings in a post-Newtonian series, following the methods
used for studying post-Newtonian Maclaurin spheroids \citep{Chandrasekhar67,Bardeen71,Petroff03b}, it does.
In this paper the first post-Newtonian corrections for homogeneous rings are calculated and the results
are compared to numerical ones generated using a version of the code described in \cite{AKM03b}, but modified
to rings, cf.~\cite{Meineletal08}.

\section{Basic Equations}

The matter model is that of a perfect fluid, i.e.\ the energy-momentum tensor reads
\begin{align}
 T^{ik} = pg^{ik} + \left(\mu+\frac{p}{c^2}\right)u^i u^k,
\end{align}
where $p$ is the pressure, $c$ the speed of light, $\mu c^2$ the energy density and $u^i$ the four-velocity.
We are considering homogeneous matter, by which we mean the mass-density is constant:
\begin{align}
 \mu=\text{constant}.
\end{align}
\par

\begin{figure}
 \centerline{\includegraphics{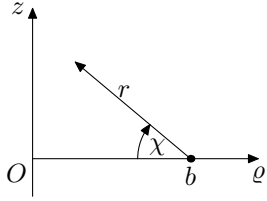}}
 \caption{A sketch providing the meaning of the coordinates $(r,\chi)$.\label{coords}}
\end{figure}

An axially symmetric, stationary spacetime containing (only) a rigidly rotating fluid can be described
using the line element
\begin{align}\label{LE}
\begin{split}
\d s^2&=\e^{2\alpha}(\d\varrho^2+\d z^2)+\varrho^2\e^{2\kappa}(\d\varphi-\omega\,\d t)^2\\
      &\quad -\e^{2\nu}c^2\,\d t^2,
\end{split}
\end{align}
where the metric functions depend only on $\varrho$ and $z$. The coordinates are uniquely
fixed by requiring that the metric functions and their first derivatives be continuous
at the fluid surface. We introduce the further coordinates $r$ and $\chi$ (see Fig.~\ref{coords})
\begin{align}\label{varrho}
 \varrho=b-r\cos\chi, \qquad z=r\sin\chi,
\end{align}
where $b$ is the `centre of mass' of the ring's cross-section, defined by
\begin{align}\label{def_b}
 b:=\frac{\iint\mu\varrho\,\d\varrho\,\d z}{\iint\mu\,\d\varrho\,\d z}
\end{align}
and we have retained $\mu$ in this and the next equation in this paper
to have expressions valid for the more general case in which $\mu$ is
not constant. The parameter $b$ is a coordinate dependent quantity and
has no particular physical significance in contrast to invariants
such as mass or angular momentum.
Inserting \eqref{varrho} in the numerator of \eqref{def_b}, one finds that
\begin{align}\label{com}
 \iint\mu r^2\cos\chi\,\d r\,\d\chi=0
\end{align}
holds. Since we are considering rigid rotation, we can introduce the constant
angular velocity $\Omega=u^\varphi/u^t$. A locally non-rotating observer would
measure the following three-velocity for a fluid element
\begin{align}\label{def_v}
 v:=\varrho(\Omega-\omega)\e^{\kappa-\nu},
\end{align}
see \citet{Bardeen70}.
\par

It is convenient to introduce the following dimensionless quantities
\begin{align}
\tilde v:=\frac{v}{c},\quad
\tilde\omega:=\frac{\omega}{\Omega},\quad
\tilde\Omega:=\frac{b\Omega}{c},\quad
\tilde p:=\frac{p}{\mu c^2}.
\end{align}
The field equations are
\begin{align}
\begin{split}
\triangle\nu&=\frac{4\pi \G\mu\e^{2\alpha}}{c^2}\left[\frac{1+\tilde v^2}{1-\tilde v^2}(1+\tilde p)+2\tilde p\right]\label{nu}\\
&\quad -\nabla\nu\cdot\nabla(\kappa+\nu)+\frac{\varrho^2\tilde\Omega^2}{2b^2}\e^{2(\kappa-\nu)}(\nabla\tilde\omega)^2,
\end{split}
\end{align}
\begin{align}
\triangle_1(\kappa+\nu)&=\frac{16\pi \G\mu}{c^2}\tilde p\e^{2\alpha}-[\nabla(\kappa+\nu)]^2,
\end{align}
\begin{align}
\begin{split}
\triangle_2\tilde\omega&=-\frac{16\pi \G\mu}{c^2}\frac{(1-\tilde\omega)(1+\tilde p)}{1-\tilde v^2}\e^{2\alpha}\\
                       &\quad  -\nabla\tilde\omega\cdot\nabla(3\kappa-\nu),
\end{split}
\end{align}
\begin{align}
\begin{split}
\triangle_{-1}\alpha&=-\frac{4\pi \G\mu}{c^2}(1+\tilde p)\e^{2\alpha}+\nabla\kappa\cdot\nabla\nu\\
                       &\quad +\frac{\varrho^2}{4b^2}\e^{2(\kappa-\nu)}(\nabla\tilde\omega)^2+
                         \frac1\varrho\frac{\partial\nu}{\partial\varrho},
\end{split}
\end{align}
where $\G$ is the gravitational constant.
Here $\nabla$ is the nabla operator in a three-space with the cylindrical coordinates
$(\varrho,\varphi,z)$ and $\triangle_m$ is a generalized Laplacian operator, which
when applied to an axially symmetric function $f=f(\varrho,z)$ gives
\begin{align}\label{Lap_m}
\triangle_m f=
\frac{\partial^2 f}{\partial\varrho^2}
+\frac{m+1}{\varrho}\frac{\partial f}{\partial\varrho}+\frac{\partial^2 f}{\partial z^2}.
\end{align}
The operator is linear and becomes the standard Laplacian for $m=0$, i.e.\ $\triangle_0=\triangle$.
The vanishing divergence of the energy-momentum tensor, $T^{ik}{}_{;k}=0$, gives
\begin{align}\label{relEuler}
 (1+\tilde p)\e^{\nu}\sqrt{1-\tilde v^2}=1-\gamma,
\end{align}
where $\gamma$ is a constant related to the relative redshift $Z_0$ via
\begin{align}\label{redshift}
Z_0=\frac{\gamma}{1-\gamma}.
\end{align}
To be more precise, this is the redshift of photons emitted from the ring's surface and observed at infinity
that carry no angular momentum.

\section{Post-Newtonian Expansion}

Let us introduce the relativistic parameter
\begin{align}
 \varepsilon:=\frac{\sqrt{\G\mu}b}{c}
\end{align}
and the following expansions
\begin{alignat}{3}
\nu&=\sum_{i=1}^\infty\nu_{2i}\varepsilon^{2i},\quad
&\kappa&=\sum_{i=1}^\infty\kappa_{2i}\varepsilon^{2i},\\
\alpha&=\sum_{i=1}^\infty\alpha_{2i}\varepsilon^{2i},\quad
&\tilde\omega&=\sum_{i=1}^\infty\tilde\omega_{2i}\varepsilon^{2i},\\
\gamma&=\sum_{i=1}^\infty\gamma_{2i}\varepsilon^{2i},\quad
&\tilde\Omega&=\sum_{i=1}^\infty\tilde\Omega_{2i-1}\varepsilon^{2i-1},\\
\tilde v&=\sum_{i=1}^\infty\tilde v_{2i-1}\varepsilon^{2i-1},\quad
&\tilde p&=\sum_{i=1}^\infty\tilde p_{2i}\varepsilon^{2i},
\end{alignat}
where the even/odd powers of $\varepsilon$ that appear in the sums come about
from stationarity (simultaneous reversal in the direction
of time $\d t\to-\d t$ and in the direction of rotation $\Omega\to -\Omega$).
The surface of the fluid, defined by vanishing pressure, will be
parameterized by the function $r=r_\text s(\chi)$, which we also
expand
\begin{align}
 r_\text s(\chi)=\sum_{i=0}^\infty r_{2i}(\chi)\varepsilon^{2i}.
\end{align}
\par

When one replaces $\G\mu/c^2$ in the field equations by $\varepsilon^2/b^2$
and expands them in $\varepsilon$, then the leading order equation comes
from \eqref{nu} and is
the familiar Laplace/Poisson equation (in a somewhat unfamiliar notation)
\begin{align}
 \triangle\nu_2^\text{o} = 0, \quad  \triangle\nu_2^\text{i} = \frac{4\pi}{b^2},
\end{align}
and the expansion of \eqref{relEuler} yields the Bernoulli equation
\begin{align}
 \tilde p_2 - \frac{\tilde v_1^2}{2} + \nu_2^\text{i} = -\gamma_2,
\end{align}
where $\tilde v_1=\varrho\tilde\Omega_1/b$ results from \eqref{def_v}
and the superscripts `i' and `o' refer to the regions inside and
outside the fluid respectively. To leading order, \eqref{com} becomes
\begin{align}
 \int_0^{2\pi} r_0^3(\chi)\cos\chi\, \d\chi=0.
\end{align}
On the Newtonian boundary $r_0(\chi)$, $\tilde p_2=0$ holds and the function $\nu_2$ and its
first derivatives must be continuous. It follows from asymptotic flatness that $\nu_2$ tends
to zero at infinity. No analytic solution for rings is known even in the Newtonian case
(described by the above equations). Instead, we solve them using an expansion
about the thin ring limit as in \cite{PH08b}. For example, the expansion for the Newtonian
surface function is
\begin{align}
 r_0(\chi)=a\left(1+\sum_{i=1}^q\sum_{k=1}^i \beta_{ik}\cos(k\chi)\sigma^i + o(\sigma^q)\right)
\end{align}
with
\begin{align}
 \sigma:=\frac{a}{b}
\end{align}
and where it will be convenient to introduce
\begin{align}
 \lambda:=\ln\frac{8}{\sigma}-2
\end{align}
for later.
The expression for $r_0$ makes use of the fact that the cross-section tends to
a circle of radius $a$ for $\sigma\to 0$, as mentioned in the introduction. The solution given
for Newtonian rings in the above paper can be used here, where the relations
\begin{align}\label{Newt_rels}
\begin{split}
 & U \equiv c^2\varepsilon^2\nu_2,
 \qquad  \Omega_\text N  \equiv cb^{-1}\varepsilon\tilde\Omega_1,
 \qquad  V_0 \equiv -c^2\varepsilon^2\gamma_2,\\
 & p_\text N  \equiv \mu c^2\varepsilon^2 \tilde p_2,
 \qquad  v_\text N  \equiv c\varepsilon\tilde v_1
\end{split}
\end{align}
have to be taken into account
and the index `N' denotes a Newtonian quantity here to distinguish
it from the relativistic quantity.
\par

Turning now to the first post-Newtonian correction, we find the
equations
\begin{align}
\triangle_1(\kappa_2+\nu_2)&=0,\\
\triangle_{-1}(\alpha_2+\nu_2)&=0,\\
\triangle_2\tilde\omega_2^\text{o}&=0,\label{genLap}\\
\triangle_2\tilde\omega_2^\text{i}&=-\frac{16\pi}{b^2},\label{omega_2^i}\\
\triangle\nu_4^\text{o}&=0,\\
\triangle\nu_4^\text{i}&=\frac{8\pi}{b^2}\left(-\nu_2^\text{i}+\tilde v_1^2+\frac{3}{2}\tilde p_2\right)
\label{nu_4^i}.
\end{align}
The first two of these equations are solved by
\begin{align}
\kappa_2=\alpha_2=-\nu_2,
\end{align}
where we remind the reader that $\nu_2$, $\tilde p_2$ and $\tilde v_1$
are already known from the Newtonian order (cf.\ \eqref{Newt_rels}) and in \eqref{nu_4^i}
$\alpha_2^\text{i}$ was replaced by $-\nu_2^\text{i}$.
Equation \eqref{relEuler} gives
\begin{align}\label{Euler_pN}
\begin{split}
&\tilde p_4+\tilde p_2\left(\nu_2^\text{i}-\frac{\tilde v_1^2}{2}\right)+ \nu_4^\text{i}\\
&\quad +\frac{\nu_2^{\text{i}2}}2-\frac12\nu_2^\text{i}\tilde v_1^2
 -\frac{\tilde v_1^4}{8}-\tilde v_1\tilde v_3
=-\gamma_4,
\end{split}
\end{align}
where
\begin{align}
\tilde v_3=\frac{\varrho}{b}\left[\tilde\Omega_3-\tilde\Omega_1(\tilde\omega_2^\text{i}+2\nu_2^\text{i})\right]
\end{align}
follows from the expansion of \eqref{def_v}. Equation \eqref{com} yields
\begin{align}\label{com_PN}
 \int_0^{2\pi}r_2(\chi)r_0(\chi)^2\cos\chi\,\d\chi=0.
\end{align}
Additionally, the transition conditions at the fluid boundary become
\begin{align}
\left(\tilde\omega_2^\text{i}-\tilde\omega_2^\text{o}\right)_{r=r_0(\chi)}&=0,\\
\left.\frac{\partial(\tilde\omega_2^\text{i}-\tilde\omega_2^\text{o})}{\partial r}\right|_{r=r_0(\chi)}&=0,\\
\left(\nu_4^\text{i}-\nu_4^\text{o}\right)_{r=r_0(\chi)}&=0,\\
\left.\frac{\partial(\nu_4^\text{i}-\nu_4^\text{o})}{\partial r}\right|_{r=r_0(\chi)}&+
r_2\left.\frac{\partial^2(\nu_2^\text{i}-\nu_2^\text{o})}{\partial r^2}\right|_{r=r_0(\chi)}=0.
\end{align}
at this order. Vanishing pressure on the surface translates to
\begin{align}\label{p4_surface}
 \tilde p_4|_{r=r_0(\chi)}+r_2(\chi)\left.\frac{\partial\tilde p_2}{\partial r}\right|_{r=r_0(\chi)}=0.
\end{align}

\section{Expansion about the Thin Ring Limit}
As with the Newtonian order, we expand the unknown functions as a power
series in $\sigma$ and a Fourier
series in $\chi$ (the sine terms do not appear as a result of reflectional
symmetry with respect to the equatorial plane):
\begin{align}
\nu_4^\text{i}&=\pi^2\sigma^4\sum_{i=0}^q\sum_{k=0}^i N^\text{i}_{ik}(y)\cos(k\chi)\sigma^i+o(\sigma^{q+4}),\\
\nu_4^\text{o}&=\pi^2\sigma^4\sum_{l=1}^{q+1}\sum_{i=l-1}^q F_{li}I_l(y,\chi)a^{2l-1}\sigma^{i-l}+o(\sigma^{q+4}),\label{nu_4^o_series}\\
\tilde\omega_2^\text{i}&=\pi\sigma^2\sum_{i=0}^q\sum_{k=0}^i w^\text{i}_{ik}(y)\cos(k\chi)\sigma^i+o(\sigma^{q+2}),\\
\tilde\omega_2^\text{o}&=\pi\sum_{l=1}^{q+1}\sum_{i=l-1}^q G_{li}K_l(y,\chi)a^{2l+1}\sigma^{i-l}+o(\sigma^{q+2})
       \label{omega^o_series}\\
\tilde p_4&=\pi^2\sigma^4\sum_{i=0}^q\sum_{k=0}^i q_{ik}(y)\cos(k\chi)\sigma^i+o(\sigma^{q+4}),\\
\tilde\Omega_3&=\pi^{3/2}\sigma^2\sum_{i=1}^{q+1} L_i\sigma^i+o(\sigma^{q+3}),\\
\gamma_4&=\pi^2\sigma^4\sum_{i=0}^{q} g_i\sigma^i+o(\sigma^{q+4}),\\
\begin{split}\label{r_2_sum}
\frac{r_2}a&\equiv y_2=\pi\sigma^2 d+\pi\sigma^2\sum_{i=1}^q\sum_{k=1}^i\kappa_{ik}\cos(k\chi)\sigma^i\\
 &\quad+o(\sigma^{q+2}),
\end{split}
\end{align}
where
\begin{align}
y:=\frac{r}{a}
\end{align}
and the parameter $d$ introduced in \eqref{r_2_sum} represents the post-Newtonian
contribution to the radius of the ring to leading order in $\sigma$. The relevance
of this parameter will be discussed below. An important property of the Fourier
series, is that they terminate due to the expansion in $\sigma$.
The functions $I_l(y,\chi)$ form an axially symmetric set of solutions to Laplace's equation, regular everywhere except
at $r=0$ and vanishing at infinity,
see \cite{PH08b}. The set $K_l(y,\chi)$ is the analogue for the generalized Laplace's equation
\eqref{genLap}, that can be found using separation of variables in toroidal coordinates $(\eta,\xi,\varphi)$.
These are defined by
\begin{align}
 \varrho &= \frac{b\sinh\xi}{\cosh\xi-\cos\eta},\\
 z       &= \frac{b\sin\eta}{\cosh\xi-\cos\eta}.
\end{align}
The set $K_l$ can be defined by
\begin{align}
 K_1&:=\frac{\pi}{\sqrt{2}b^3}\frac{(\cosh\xi-\cos\eta)^{3/2}}{\sinh\xi}P^1_{-1/2}(\cosh\xi),\\
 K_l&:=\left(-\frac{1}{b}\frac{\d}{\d b}\right)^{l-1}K_1\intertext{with}
 \frac\d{\d b}&:=\frac\partial{\partial b} + \cos\chi\frac\partial{\partial r}-\frac{\sin\chi}{r}\frac\partial{\partial\chi}.
\end{align}
On the axis, they have the simple form
\begin{align}
 K_l(r)=-\frac{\pi(2l-1)!!}{4r^{2l+1}}\quad\text{(on axis)}.
\end{align}
The idea now is to expand these functions with respect to $\sigma$ and
then proceed to solve the field equations iteratively. We demonstrate
the technique with the leading order in $\sigma$.

\section{The Solution to the Equations}

The post-Newtonian correction to the surface to leading order
in $\sigma$ is
\begin{align}
 y_2(\chi)=\pi\sigma^2[d+o(\sigma^0)].
\end{align}
As with $y_0$, this function is also independent of $\chi$ to leading order and
fulfils \eqref{com_PN}. Inside the ring, \eqref{omega_2^i}
becomes
\begin{align}
 \frac{\d^2w^\text{i}_{00}}{\d y^2}+\frac1{y}\frac{\d w^\text{i}_{00}}{\d y}+16=0,
\end{align}
whose regular solution is
\begin{align}
 w^\text{i}_{00}=-4y^2+c_1.
\end{align}
The transition conditions are satisfied for
\begin{align}
 c_1=8\lambda+4
\end{align}
and where the constant $G_{10}$ from \eqref{omega^o_series} is
\begin{align}
G_{10}=-16.
\end{align}
\par

The generalized Poisson equation \eqref{nu_4^i} becomes
\begin{align}
 \frac{\d^2N^\text{i}_{00}}{\d y^2}+\frac1{y}\frac{\d N^\text{i}_{00}}{\d y}+20y^2-24\lambda-58=0
\end{align}
and we can immediately write down the regular solution
\begin{align}
  N^\text{i}_{00}=-\frac{5}{4}y^4+\left(6\lambda+\frac{29}2\right)y^2+c_2.
\end{align}
Here, the transition conditions lead to
\begin{align}
 c_2= -12\lambda^2-54\lambda-4\lambda d-8d-\frac{245}4
\end{align}
and for $F_{10}$ from \eqref{nu_4^o_series}
\begin{align}
 F_{10}=-12\lambda-4d-24.
\end{align}
\par

The equilibrium condition \eqref{Euler_pN} gives an expression for the
pressure coefficient $q_{00}$ that can be combined with \eqref{p4_surface} to give
\begin{align}
  q_{00}=\frac{7}{4}y^4-(4\lambda+14)y^2+4\lambda+\frac{49}4+2d,
\end{align}
thus completing the solution to leading order in $\sigma$.
\par

In general, stationary and axially symmetric fluids with a given equation of state
are characterized by two parameters, e.g.\ gravitational mass and angular momentum.
The corresponding post-Newtonian expansions
contain two additional free parameters, which arise because of the freedom one has in
choosing which two quantities contain no post-Newtonian contribution, see \cite{Bardeen71} and discussion
therein. We too would obtain
two free parameters, but have already chosen to have $b$ remain unchanged to first
order in $\varepsilon$ and thus only have $d$ as a free parameter in the solution.%
\footnote{Note that to leading order in $\sigma$, one only has the freedom
to fix a single parameter. The parameter $b$ does not appear to this order
and we retain the freedom to choose $d$.}
\par

The leading order in $\sigma$ is equivalent to an infinite
cylinder, as in the Newtonian case, see \cite{Ostriker64,Ostriker64b}.
In \cite{Bicaketal04}, relativistic, homogeneous cylinders and their post-Newtonian
expansion are investigated. In that paper, there is no post-Newtonian contribution
to the central pressure, which corresponds to choosing $d=-2\lambda-49/8$ in our case.
In order to compare our results to theirs, we introduce the proper radial coordinate
and circumferential radial coordinate
\begin{align}
 y_\text{p} &:= \frac{1}{a}\int_0^r\sqrt{g_{rr}}\,\d r'=\int_0^y\e^\alpha\,\d y',\\
 y_\text{c} &:= \frac{1}{2\pi a}\int_0^{2\pi}\sqrt{g_{\chi\chi}}\, \d\chi'=y\e^\alpha.
\end{align}
The values of these coordinates on the cylinder's surface (using the choice for $d$ given above)
are found to be
\begin{align}\begin{aligned}
 y_\text{p} = 1-\frac{35}{24}\pi\sigma^2\varepsilon^2[1+o(\sigma^0)]+O(\varepsilon^4),\\
 y_\text{c} = 1-\frac{17}{8}\pi\sigma^2\varepsilon^2[1+o(\sigma^0)]+O(\varepsilon^4),
\end{aligned}\quad\text{(on surface)}\end{align}
in agreement with (6.10) and (6.11) in \cite{Bicaketal04}. The expressions
for the pressure, e.g.\ as a function of proper radial coordinate, can also be shown to agree.
\par

The expansion to higher orders in $\sigma$ presents no particular difficulties and
we have carried it out up to seventh order. The resulting coefficients can be found
in Appendix~\ref{appendix_coefficients} up to fourth order, but the results presented
in the next section are given up to seventh order.

\section{Results}
In the following it will be convenient to use dimensionless quantities, denoted by a bar:
\begin{alignat}{3}
\frac{\bar{\Omega}^2}{\Omega^2}&=\frac{1}{\G\mu},\qquad&
\frac{\bar{M}}M&=\frac{\bar{M}_\text{B}}{M_\text{B}}=\frac{\G^{3/2}\mu^{1/2}}{c^3},\\
\frac{\bar{J}}{J}&=\frac{\G^2\mu}{c^5},\qquad&
\frac{\bar{b}}{b}&=
\frac{\bar{\varrho}}{\varrho}=
\frac{\bar{z}}{z}=
\frac{\G^{1/2}\mu^{1/2}}c,
\end{alignat}
where $M$ is the gravitational mass, $M_\text{B}$ the baryonic mass and $J$
the angular momentum.
In particular
\begin{align}
 \bar{b}=\frac{\sqrt{\G\mu}b}c=\varepsilon
\end{align}
holds.
\begin{table*}
\centering \caption{\label{different_d} Comparison between the
Newtonian and first post-Newtonian (pN) results up to seventh order in $\sigma$.
The relative errors are computed by comparing with a highly accurate numerical
solution with the prescribed values $\varrho_\text{i}/\varrho_\text{o}=0.7$
and $Z_0=0.05$. For the definition of $d_\Omega$, $d_J$ and $d_M$,
  see \eqref{def_d}.}
$
\begin{array}{D{=}{=}{12}lllll}\toprule
\multicolumn{1}{c}{\text{Numerical value}}     & \multicolumn{5}{c}{\text{Relative error}} \\  \cmidrule{2-6}
\multicolumn{1}{c}{\text{Relativistic}}       & \text{Newtonian}             & \text{pN:\,}d=0               & \text{pN:\,}d=d_\Omega        & \text{pN:\,}d=d_J             & \text{pN:\,}d=d_M\\ \midrule
\bar{\Omega}=4.9108\times10^{-1}             & -2\times10^{-2}           & -1\times10^{-3}           & -3\times10^{-4}           & \phantom{+}3\times10^{-3} & \phantom{+}2\times10^{-3}\\
\bar{J}=2.3168\times10^{-4}                  & \phantom{+}1\times10^{-1} & \phantom{+}1\times10^{-2} & \phantom{+}8\times10^{-3} & -8\times10^{-3}           & -2\times10^{-3}          \\
\bar{M}=7.9661\times10^{-3}                  & \phantom{+}9\times10^{-2} & \phantom{+}1\times10^{-2} & \phantom{+}7\times10^{-3} & -5\times10^{-3}           & -1\times10^{-3}          \\
\bar{M}_\text{B}=8.0842\times10^{-3}         & \phantom{+}7\times10^{-2} & \phantom{+}1\times10^{-2} & \phantom{+}5\times10^{-3} & -4\times10^{-3}           & -1\times10^{-3}          \\
\bottomrule
\end{array}
$
\end{table*}
In Table~\ref{different_d}, the Newtonian and first post-Newtonian results
for the ring with the prescribed values $\varrho_\text{i}/\varrho_\text{o}=0.7$ and $Z_0=0.05$
are compared with a highly accurate numerical solution in full General Relativity.
In Newtonian theory, there is (strictly speaking) no redshift, but the leading term,
$Z_0=-V_0/c^2$, contains only the Newtonian quantity $V_0$ and
the speed of light, showing the relativistic origin. This equation explains what is meant by the
`redshift' of a Newtonian configuration. The free parameter $d$ is chosen in a number of different ways in the table.
For one, we simply choose $d=0$. Alternatively, we expand a chosen quantity
\begin{align}
 \begin{split}
 \bar\Omega&=\bar\Omega_\text{N}+\bar\Omega_\text{pN}\varepsilon^2+O(\varepsilon^4),\\
     \bar J&=    \bar J_\text{N}+    \bar J_\text{pN}\varepsilon^2+O(\varepsilon^4),\\
     \bar M&=    \bar M_\text{N}+    \bar M_\text{pN}\varepsilon^2+O(\varepsilon^4),
 \end{split}
\end{align}
and define $d$ so that there is no post-Newtonian contribution%
\footnote{Nonetheless even for this
quantity the values in the Newtonian and post-Newtonian column are different, because $\varrho_\text{i}/\varrho_\text{o}=0.7$ and
$Z_0=0.05$ are prescribed, and not the chosen quantity.},
i.e.\
\begin{align}\label{def_d}
 \begin{split}
  \bar\Omega_\text{pN}(d=d_\Omega)&=0,\\
  \bar     J_\text{pN}(d=d_J)&=0,\\
  \bar     M_\text{pN}(d=d_M)&=0.
 \end{split}
\end{align}
The solutions to these equations when expressed as series in $\sigma$ are
\begin{align}
 \begin{split}
  d_\Omega&=-\frac{144\lambda^2+240\lambda+119}{24(4\lambda+1)}\\
    &\quad  +\frac{6912\lambda^3+864\lambda^2-9684\lambda-5903}{1152(4\lambda+1)^2}\sigma^2
            +o(\sigma^3)\\
  d_J&=     -\frac{624\lambda^2+3120\lambda+2009}{24(12\lambda+7)}\\
   &\quad +\frac{202752\lambda^3+1487136\lambda^2+1765236\lambda+549479}{1152(12\lambda+7)^2}\sigma^2\\
   &\quad +o(\sigma^3)\\
  d_M&=-3(\lambda+2)+\frac{15}{64}\sigma^2+o(\sigma^3).
 \end{split}
\end{align}
\par

Throughout Table~\ref{different_d}, the post-Newtonian expressions yield a clear improvement compared to the Newtonian ones.
The deviations from the accurate numerical values have two origins: Terms in $\sigma$ and terms in $\varepsilon$ are missing.
\begin{table*}
 \centering \caption{\label{Omega_pN_q}
 Comparison between the Newtonian and first post-Newtonian (pN) results for the
 dimensionless angular velocity $\bar\Omega$ for different orders $q$ in $\sigma$.
 The prescribed parameters for the ring are $\varrho_\text{i}/\varrho_\text{o}=0.7$
 and $Z_0=0.05$. The numerical calculations show that the value for $\bar\Omega$
 for the relativistic ring is $\bar{\Omega}_\text{num}=0.49108$ (cf.\ Table~\ref{different_d})
 and for the Newtonian one is $\bar{\Omega}_\text{N,num}=0.48109$
 (see Table~1 in \citet{AKM03c}). For the definition of $d_\Omega$, $d_J$ and $d_M$,
  see \eqref{def_d}.}
$
\begin{array}{cccccc}\toprule
q  & \text{Newtonian} & \text{pN:\,}d=0 & \text{pN:\,}d=d_\Omega & \text{pN:\,}d=d_J & \text{pN:\,}d=d_M \\ \midrule
1  & 0.50085      & 0.51126    & 0.51192           & 0.51364      & 0.51301\\
3  & 0.48274      & 0.49206    & 0.49264           & 0.49423      & 0.49369 \\
5  & 0.48135      & 0.49059    & 0.49116           & 0.49277      & 0.49223 \\
7  & 0.48114      & 0.49035    & 0.49092           & 0.49255      & 0.49200 \\
20 & 0.48109      & \text{---} & \text{---}        & \text{---}   & \text{---} \\
\bottomrule
\end{array}
$
\end{table*}
Table~\ref{Omega_pN_q} demonstrates that for rings with $\varrho_\text{i}/\varrho_\text{o}\gtrsim 0.7$, $q=7$ is sufficient
for relative errors smaller than $10^{-4}$ in the Newtonian case. One can estimate that the error due to terminating
the series in $\sigma$ at the order $q$ results in a relative error of about $(\sigma|\ln\sigma|)^q$, where
it is convenient to use
\begin{align}\label{sig_approx}
\sigma\sim \frac12\left(1-\frac{\varrho_\text{i}} {\varrho_\text{o}}\right).
\end{align}
 If the relativistic effects are large in comparison
(as is the case in Table~\ref{Omega_pN_q} although $Z_0$ is only $0.05$), then truncating in $\sigma$ is
negligible. A rough estimate for $\varepsilon$ can be found by taking the leading term for the redshift in
both $\varepsilon$ and $\sigma$, inserting \eqref{sig_approx} and approximating slightly
to find
\begin{align}
\varepsilon\sim\frac{1}{1-\varrho_\text{i}/\varrho_\text{o}}\sqrt{\frac{Z_0}{6-2\ln(1-\varrho_\text{i}/\varrho_\text{o})}}.
\end{align}
The relative factor between the Newtonian and post-Newtonian contribution to $\Omega$ is about
$5|\ln\sigma|(\sigma\varepsilon)^2\sim10^{-2}$ for the values from Table~\ref{Omega_pN_q}.
\par

Tables~\ref{different_d_95} and \ref{Omega_pN_q_95} present analogous results for $\varrho_\text{i}/\varrho_\text{o}=0.95$
and $Z_0=10^{-3}$. The values of each expansion parameter (i.e.\ $\sigma$ and $\varepsilon$)
are small enough that extremely high accuracy is observed. This also permits us
to see how the results improve as one proceeds to the post-Newtonian level and with
increasing orders in $\sigma$. Tables~\ref{different_d} and \ref{different_d_95}
suggest that the choice $d=d_x$ is particularly accurate if the quantity
$x$ is prescribed.
\begin{table*}
\centering \caption{\label{different_d_95} Comparison between the
Newtonian and first post-Newtonian (pN) results up to seventh order in $\sigma$.
The relative errors are computed by comparing with a highly accurate numerical
solution with the prescribed values $\varrho_\text{i}/\varrho_\text{o}=0.95$
and $Z_0=10^{-3}$. For the definition of $d_\Omega$, $d_J$ and $d_M$,
  see \eqref{def_d}.}
$
\begin{array}{D{=}{=}{15}lllll}\toprule
\multicolumn{1}{c}{\text{Numerical value}}     & \multicolumn{5}{c}{\text{Relative error}} \\  \cmidrule{2-6}
\multicolumn{1}{c}{\text{Relativistic}}   & \text{Newtonian}             & \text{pN:\,}d=0               & \text{pN:\,}d=d_\Omega        & \text{pN:\,}d=d_J             & \text{pN:\,}d=d_M\\ \midrule
\bar{\Omega}=9.6226144\times10^{-2}       & -5\times10^{-4}           & -1\times10^{-6}           & -3\times10^{-7}           & \phantom{+}8\times10^{-7} & \phantom{+}7\times10^{-7}\\
\bar{J}=2.9191516\times10^{-7}            & \phantom{+}3\times10^{-3} & \phantom{+}7\times10^{-6} & \phantom{+}4\times10^{-6} & -2\times10^{-6}           & -2\times10^{-6}          \\
\bar{M}=8.5891100\times10^{-5}            & \phantom{+}1\times10^{-3} & \phantom{+}5\times10^{-6} & \phantom{+}2\times10^{-6} & -1\times10^{-6}           & -7\times10^{-7}          \\
\bar{M}_\text{B}=8.5914515\times10^{-5}   & \phantom{+}1\times10^{-3} & \phantom{+}4\times10^{-6} & \phantom{+}2\times10^{-6} & -9\times10^{-7}           & -8\times10^{-7}          \\
\bottomrule
\end{array}
$
\end{table*}
\begin{table*}
 \centering \caption{\label{Omega_pN_q_95}
 Comparison between the Newtonian and first post-Newtonian (pN) results for the
 dimensionless angular velocity $\bar\Omega$ for different orders $q$ in $\sigma$.
 The prescribed parameters for the ring are $\varrho_\text{i}/\varrho_\text{o}=0.95$
 and $Z_0=10^{-3}$. The numerical calculations show that the value for $\bar\Omega$
 for the relativistic ring is $\bar{\Omega}_\text{num}=0.096226144$ (cf.\ Table~\ref{different_d_95})
 and for the Newtonian one is $\bar{\Omega}_\text{N,num}=0.096177081$. For the definition of $d_\Omega$, $d_J$ and $d_M$,
  see \eqref{def_d}.}
$
\begin{array}{cccccc}\toprule
q  & \text{Newtonian} & \text{pN:\,}d=0 & \text{pN:\,}d=d_\Omega & \text{pN:\,}d=d_J & \text{pN:\,}d=d_M \\ \midrule
1  & 0.096333800      & 0.096383025     & 0.096383103            & 0.096383208       & 0.096383200\\
3  & 0.096177653      & 0.096226613     & 0.096226690            & 0.096226795       & 0.096226786\\
5  & 0.096177085      & 0.096226044     & 0.096226121            & 0.096226226       & 0.096226217\\
7  & 0.096177081      & 0.096226040     & 0.096226117            & 0.096226222       & 0.096226214\\
20 & 0.096177081      & \text{---} & \text{---}        & \text{---}   & \text{---} \\
\bottomrule
\end{array}
$
\end{table*}
\par

\begin{figure*}
 \centerline{\includegraphics{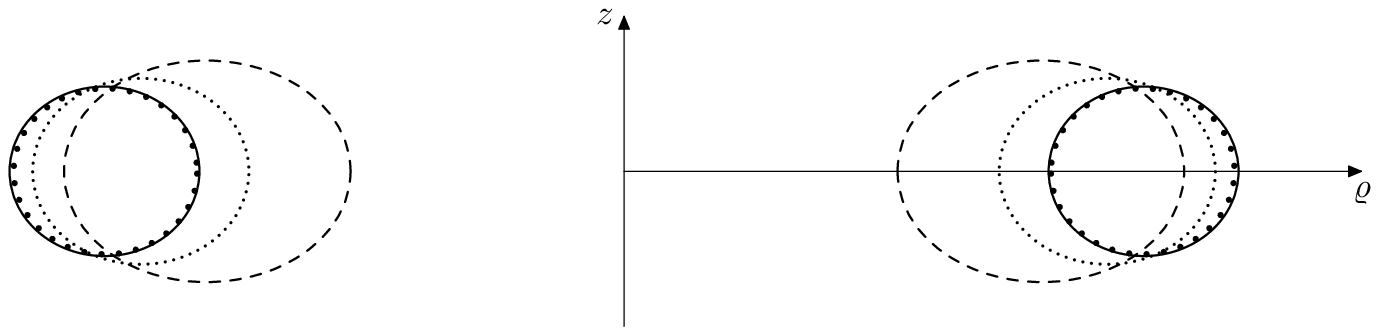}}
 \caption{The cross-section of the ring with $\bar{M}=0.05$ and $Z_0=0.2$. The solid line is the highly accurate
          numerical result. The large dots, which coincide with it almost exactly, represent the post-Newtonian result
          with $d=d_M$ (cf.\ \eqref{def_d}). The thin dotted line is the post-Newtonian result with $d=0$ and the dashed line is the
          Newtonian result. All analytic curves were generated with $q=7$.\label{cross-section}}
\end{figure*}
Figure~\ref{cross-section} provides an example of the meridional cross-section of the surface shape of
a ring. Both post-Newtonian results given there can be seen to constitute a marked improvement as
compared to the Newtonian one. Moreover, we find that the choice $d=d_M$ gives a significantly better
result than $d=0$, which is also the case for most values in Tables~\ref{different_d} and \ref{different_d_95}.
\par

\begin{figure}
 \centerline{\includegraphics{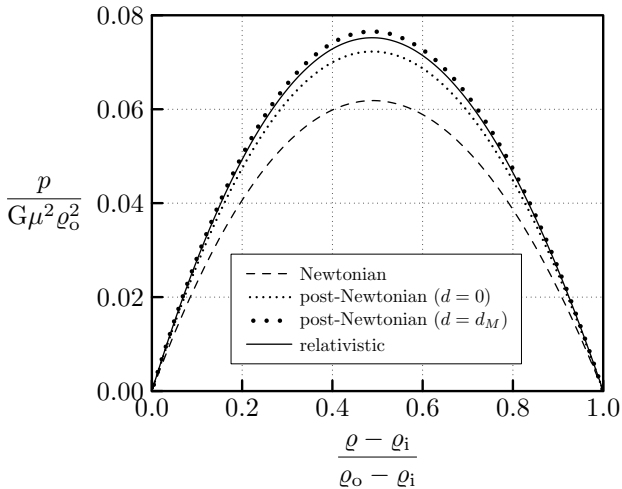}}
 \caption{The pressure profile of the ring with $A=0.7$ and $Z_0=0.1$ in the equatorial plane. For the definition of $d_\Omega$, $d_J$ and $d_M$,
  see \eqref{def_d}.\label{pressure}}
\end{figure}
A similar behaviour can be found in Fig.~\ref{pressure}, where the pressure in the equatorial plane is plotted.
Again the post-Newtonian results demonstrate an improvement as compared to the Newtonian, and here too the
choice $d=d_M$ leads to better results than $d=0$.
\par

\begin{figure}
 \centerline{\includegraphics{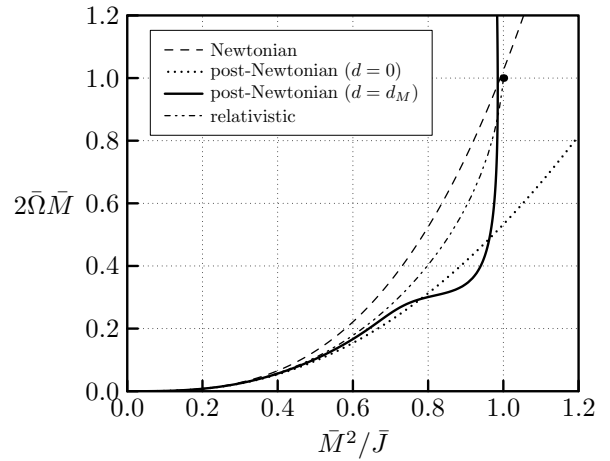}}
 \caption{The sequence of rings with $\varrho_\text{i}/\varrho_\text{o}=0.95$ for $q=7$. For the definition of $d_\Omega$, $d_J$ and $d_M$, see \eqref{def_d}.\label{2OmM_M2J_A95}}
\end{figure}
\begin{figure}
 \centerline{\includegraphics{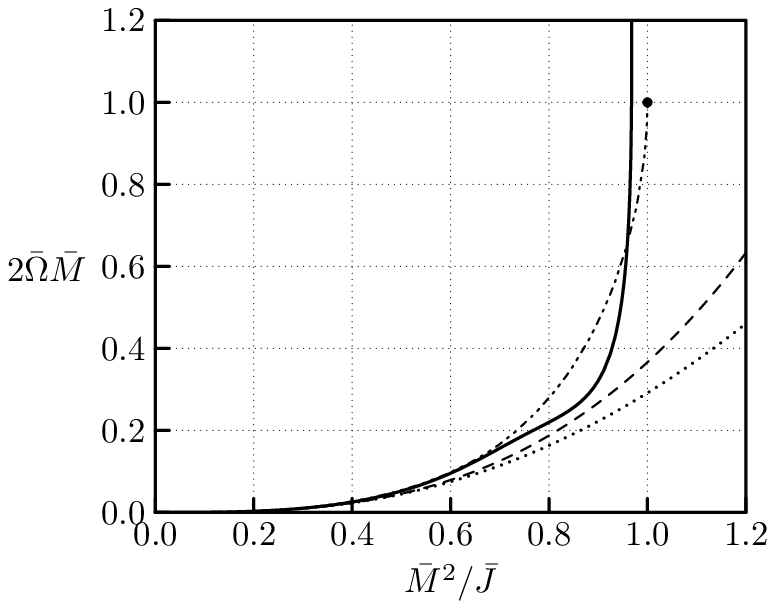}}
 \caption{The sequence of rings with $\varrho_\text{i}/\varrho_\text{o}=0.7$ for $q=7$. The line types
          are the same as in Fig.~\ref{2OmM_M2J_A95}.\label{2OmM_M2J_A70}}
\end{figure}
\begin{figure}
 \centerline{\includegraphics{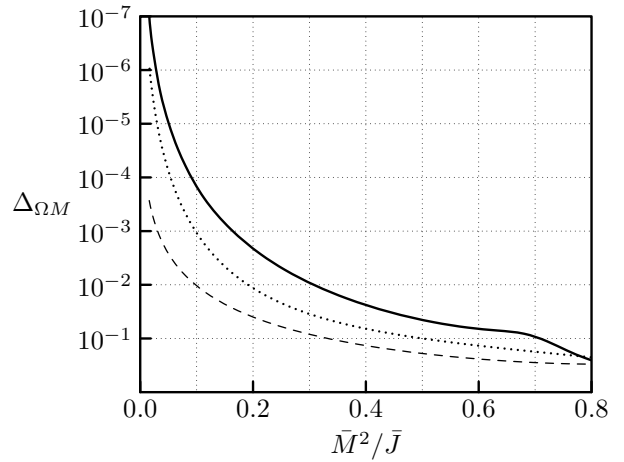}}
 \caption{The relative errors for the sequence of rings corresponding to Fig.~\ref{2OmM_M2J_A95}
  ($\varrho_\text{i}/\varrho_\text{o}=0.95$) determined by comparing them to the relativistic curve. The line types
          are the same as in Fig.~\ref{2OmM_M2J_A95}. \label{log_A95}}
\end{figure}
\begin{figure}
 \centerline{\includegraphics{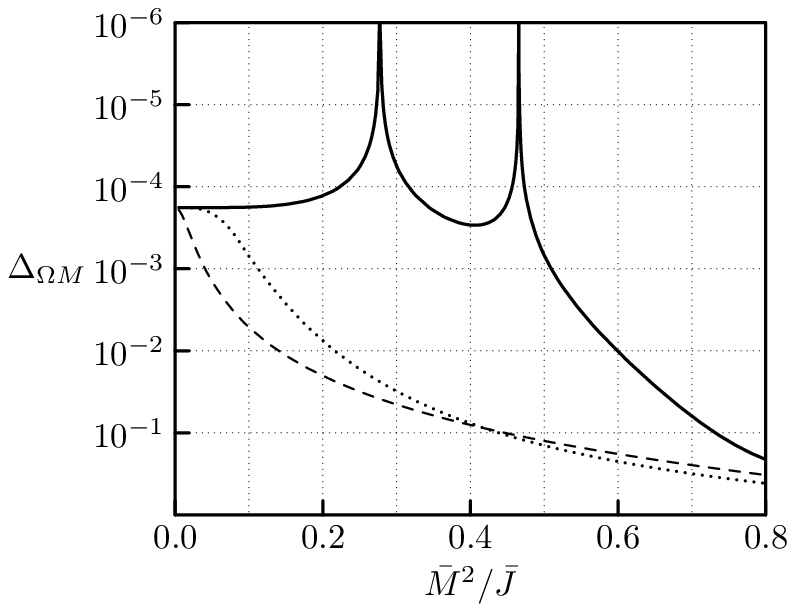}}
 \caption{The relative errors for the sequence of rings corresponding to Fig.~\ref{2OmM_M2J_A70}
  ($\varrho_\text{i}/\varrho_\text{o}=0.7$) determined by comparing them to the relativistic curve. The line types
          are the same as in Fig.~\ref{2OmM_M2J_A95}. (The two singularities to be seen in the post-Newtonian
          curve with $d=d_M$ reflect nothing more than the fact that it crosses the correct relativistic curve
          twice.)\label{log_A70}}
\end{figure}

The quality of the approximations can be seen for sequences of rings with the prescribed radius
ratios $\varrho_\text{i}/\varrho_\text{o}=0.95$ (Fig.~\ref{2OmM_M2J_A95}) and
$\varrho_\text{i}/\varrho_\text{o}=0.7$ (Fig.~\ref{2OmM_M2J_A70}). The relativistic curve
`begins' in the Newtonian limit at the origin and `ends' in an extreme Kerr Black Hole at
the point $(1,1)$. The parametric transition of rings to Black Holes was first studied
in \cite{AKM03c} and had already been considered analytically for discs \citep{NM93,Meinel02}.
It is not surprising that the accuracy of the Newtonian and post-Newtonian curves
in the vicinity of this ultra-relativistic point (at which $Z_0\to\infty$) is so poor.%
\footnote{It is a mere coincidence that some of the curves almost pass through the point $(1,1)$.}
In contrast, the highly relativistic regime
can be probed quite accurately using very high orders of the post-Newtonian approximation,
at least for the disc of dust \citep{BW69,BW71,PM01}. In the Newtonian regime, $\bar M^2/\bar J\to 0$,
the curves become indistinguishable, as they must. Naturally, we require that there exist some neighbourhood
of the Newtonian limit in which the post-Newtonian approximation provides
an improvement compared with the Newtonian results. In order to check this, we have plotted the
relative errors $\Delta_{\Omega M}=1-(\Omega M)/(\Omega M)_\text{relativistic}$ in Figs~\ref{log_A95} and
\ref{log_A70}. The errors corresponding to the sequence $\varrho_\text{i}/\varrho_\text{o}=0.95$ shown in
Fig.~\ref{log_A95} demonstrate the required behaviour. It is particularly apparent that $d=d_M$ produces
better results than $d=0$. For $\varrho_\text{i}/\varrho_\text{o}=0.7$, the Newtonian limit shown in
Fig.~\ref{log_A70} seems to be in contradiction to the requirement: the post-Newtonian approximations
in this plot provide no improvement. For this relatively small radius ratio, the truncation of the series
at the order $q=7$ provides the limiting value to the accuracy of $\sim 10^{-4}$ as discussed above. On the other
hand, the limitation in the relative error of $\sim 10^{-8}$ due to $q=7$ for $\varrho_\text{i}/\varrho_\text{o}=0.95$
is beyond the accuracy shown in Fig.~\ref{log_A95}.
\par

We have seen, that as with spheroidal bodies, the post-Newtonian approximation for toroidal
ones can be used to great advantage. It is noteworthy that this work is possible without
the existence of a known exact ring solution, even in Newtonian theory. It was therefore necessary
to expand about the thin ring limit, which we were able to do up to high orders, making
the results applicable even to fairly `thick' rings.

\section*{Acknowledgments}
   We are grateful to Reinhard Meinel for helpful discussions.
   SH would in particular like to thank him for the enjoyable and fruitful years of collaboration.
   This research was funded in part by the Deutsche Forschungsgemeinschaft
   (SFB/TR7--B1).

 \bibliographystyle{mn2e}
 \bibliography{Reflink}


\appendix
\section{The coefficients}\label{appendix_coefficients}
In the Tables~\ref{L_table}--\ref{G_table} the post-Newtonian coefficients up to the order $q=4$
are presented.
\begin{table*}
 \centering
 \renewcommand{\arraystretch}{1.8}
\caption{Coefficients $L_i$ up to the order $q=4$.\label{L_table}}
\begin{tabular}{l*{1}{>{\normalsize\PBS\raggedright}p{0.8\textwidth}}}
\toprule
 $i$ & $L_i$\\
\midrule
$1$ & $\frac{1}{48\sqrt {4 \lambda+3}}\left(144 {\lambda}^{2}+96 \lambda d+240 \lambda+24 d+119
\right)$\\
$2$ & $0$\\
$3$ & $-\frac{1}{1152\left( 4 \lambda+3 \right) ^{3/2}}
\left(3744 \lambda d+1728 {\lambda}^{2}d+1439
+7236 \lambda+6048 {\lambda}^{3}+2076 d+11448 {\lambda}^{2}\right)$\\
$4$ & $0$\\
\bottomrule
\end{tabular}
\end{table*}
\begin{table*}
 \centering
 \renewcommand{\arraystretch}{1.8}
  \caption{Coefficients $\kappa_{ik}$ up to the order $q=4$.}
\begin{tabular}{ll*{1}{>{\normalsize\PBS\raggedright}p{0.8\textwidth}}}
\toprule
$i$ & $k$ & $\kappa_{ik}$\\
\midrule
$1$ & $1$ &$0
$\\$2$ & $1$ &$0
$\\$2$ & $2$ &$\frac52 {\lambda}^{2}+{\frac {15}{8}} d\lambda+{\frac {107}{48}} 
\lambda+{\frac {671}{1152}}+{\frac {15}{32}} d
$\\$3$ & $1$ &$0
$\\$3$ & $2$ &$0
$\\$3$ & $3$ &$-{\frac {9}{32}} {\lambda}^{2}+{\frac {5}{32}} d\lambda-{\frac {1937
}{3072}} \lambda-{\frac {19825}{73728}}-{\frac {65}{768}} d
$\\$4$ & $1$ &$0
$\\$4$ & $2$ &${\frac {25}{8}} {\lambda}^{3}+{\frac {39}{16}} {\lambda}^{2}+{\frac 
{25}{8}} d{\lambda}^{2}-{\frac {515}{1152}} \lambda+{\frac {315}{128
}} d\lambda-{\frac {2093}{6912}}-{\frac {1115}{9216}} d
$\\$4$ & $3$ &$0
$\\$4$ & $4$ &${\frac {75}{32}} {\lambda}^{3}+{\frac {8111}{2304}} {\lambda}^{2}+{
\frac {375}{256}} d{\lambda}^{2}+{\frac {2725}{2304}} d\lambda+{
\frac {328073}{184320}} \lambda+{\frac {4190617}{13271040}}+{\frac {
5885}{55296}} d
$\\\bottomrule
\end{tabular}

\end{table*}
\begin{table*}
 \centering
 \renewcommand{\arraystretch}{1.8}\renewcommand{\baselinestretch}{1.4}
  \caption{Coefficients $N^\text{i}_{ik}(y)$ up to the order $q=4$.}
\begin{tabular}{ll*{1}{>{\normalsize\PBS\raggedright}p{0.8\textwidth}}}
\toprule
$i$ & $k$ & $N^\text{i}_{ik}$\\
\midrule
$0$ & $0$ &$-12 {\lambda}^{2}-54 \lambda-4 d\lambda-{\frac {245}{4}}-8 d+
 \left( 6 \lambda+{\frac {29}{2}} \right) {y}^{2}-\frac54 {y}^{4}
$\\$1$ & $0$ &$0
$\\$1$ & $1$ &$ \left( -6 {\lambda}^{2}-16 \lambda-2 d\lambda-{\frac {47}{4}}-d
 \right) y+ \left( \frac12 \lambda+{\frac {7}{2}} \right) {y}^{3}-{\frac 
{5}{12}} {y}^{5}
$\\$2$ & $0$ &$\frac54 {\lambda}^{2}+\frac12 d\lambda+{\frac {133}{24}} \lambda+\frac34 d+{
\frac {1123}{192}}+ \left( -\frac32 {\lambda}^{2}-6 \lambda-\frac12 d
\lambda-\frac14 d-{\frac {461}{96}} \right) {y}^{2}+ \left( {\frac {19}{
16}} \lambda+{\frac {127}{64}} \right) {y}^{4}-{\frac {5}{32}} {y}^{
6}
$\\$2$ & $1$ &$0
$\\$2$ & $2$ &$ \left( -\frac{17}2 {\lambda}^{2}-2 d\lambda-{\frac {59}{3}} \lambda-{
\frac {11159}{1152}}-{\frac {7}{24}} d \right) {y}^{2}+ \left( {
\frac {67}{24}} \lambda+{\frac {217}{72}} \right) {y}^{4}-{\frac {65}
{384}} {y}^{6}
$\\$3$ & $0$ &$0
$\\$3$ & $1$ &$ \left( {\frac {85}{16}} {\lambda}^{2}+{\frac {13}{8}} d\lambda+{
\frac {343}{32}} \lambda+{\frac {11987}{2304}}+{\frac {91}{96}} d
 \right) y+ \left( -{\frac {13}{4}} {\lambda}^{2}-{\frac {7}{8}} d
\lambda-{\frac {1727}{192}} \lambda-{\frac {25}{96}} d-{\frac {24167
}{4608}} \right) {y}^{3}+ \left( {\frac {257}{192}} \lambda+{\frac {
1307}{576}} \right) {y}^{5}-{\frac {25}{128}} {y}^{7}
$\\$3$ & $2$ &$0
$\\$3$ & $3$ &$ \left( -{\frac {97}{96}} {\lambda}^{2}-{\frac {15}{32}} d\lambda-{
\frac {14185}{4608}} \lambda-{\frac {50857}{36864}}+{\frac {5}{768}}
 d \right) {y}^{3}+ \left( {\frac {515}{768}} \lambda+{\frac {16465}
{18432}} \right) {y}^{5}-{\frac {7}{96}} {y}^{7}
$\\$4$ & $0$ &$-{\frac {125}{16}} {\lambda}^{4}-{\frac {75}{32}} d{\lambda}^{3}-{
\frac {1745}{64}} {\lambda}^{3}-{\frac {47291}{1536}} {\lambda}^{2}-
{\frac {55}{16}} d{\lambda}^{2}-{\frac {1097}{1536}} d\lambda-{
\frac {623875}{55296}} \lambda+{\frac {1633}{9216}} d-{\frac {118327
}{221184}}+ \left( {\frac {75}{64}} {\lambda}^{3}+{\frac {405}{256}}
 {\lambda}^{2}+{\frac {3767}{3072}} \lambda+{\frac {13}{32}} d
\lambda+{\frac {91}{384}} d+{\frac {9767}{12288}} \right) {y}^{2}+
 \left( -{\frac {39}{32}} {\lambda}^{2}-{\frac {873}{256}} \lambda-{
\frac {21}{64}} d\lambda-{\frac {25}{256}} d-{\frac {26059}{12288}}
 \right) {y}^{4}+ \left( {\frac {135}{256}} \lambda+{\frac {1415}{
1536}} \right) {y}^{6}-{\frac {675}{8192}} {y}^{8}
$\\$4$ & $1$ &$0
$\\$4$ & $2$ &$ \left( -{\frac {55}{8}} {\lambda}^{3}-\frac52 d{\lambda}^{2}-{\frac {
3517}{256}} {\lambda}^{2}-{\frac {151}{128}} d\lambda-{\frac {246605
}{36864}} \lambda+{\frac {534997}{884736}}+{\frac {5329}{9216}} d
 \right) {y}^{2}+ \left( -{\frac {217}{384}} {\lambda}^{2}-{\frac {
185}{384}} d\lambda-{\frac {6933}{2048}} \lambda-{\frac {985}{9216}}
 d-{\frac {1142287}{442368}} \right) {y}^{4}+ \left( {\frac {2389}{
3072}} \lambda+{\frac {95237}{73728}} \right) {y}^{6}-{\frac {175}{
1536}} {y}^{8}
$\\$4$ & $3$ &$0
$\\$4$ & $4$ &$ \left( -{\frac {2821}{2304}} {\lambda}^{2}-{\frac {455}{2304}} d
\lambda-{\frac {418169}{184320}} \lambda+{\frac {1001}{55296}} d-{
\frac {10764521}{13271040}} \right) {y}^{4}+ \left( {\frac {2611}{9216
}} \lambda+{\frac {85229}{221184}} \right) {y}^{6}-{\frac {133}{4096}
} {y}^{8}
$\\\bottomrule
\end{tabular}

\end{table*}
\begin{table*}
 \centering
 \renewcommand{\arraystretch}{1.8}\renewcommand{\baselinestretch}{1.4}
  \caption{Coefficients $F_{li}$ up to the order $q=4$.}
\begin{tabular}{ll*{1}{>{\normalsize\PBS\raggedright}p{0.8\textwidth}}}
\toprule
$l$ & $i$ & $F_{li}$\\
\midrule
$1$ & $0$ &$-12 \lambda-24-4 d
$\\$1$ & $1$ &$0
$\\$1$ & $2$ &${\frac {15}{16}}
$\\$1$ & $3$ &$0
$\\$1$ & $4$ &$-{\frac {125}{16}} {\lambda}^{3}-{\frac {75}{32}} d{\lambda}^{2}-{
\frac {555}{32}} {\lambda}^{2}-{\frac {17357}{1536}} \lambda-{\frac 
{125}{64}} d\lambda-{\frac {115483}{55296}}-{\frac {175}{512}} d
$\\$2$ & $1$ &$\frac52 \lambda+{\frac {10}{3}}+d
$\\$2$ & $2$ &$0
$\\$2$ & $3$ &${\frac {135}{8}} {\lambda}^{2}+{\frac {1939}{64}} \lambda+{\frac {75
}{8}} d\lambda+{\frac {125}{32}} d+{\frac {18527}{1536}}
$\\$2$ & $4$ &$0
$\\$3$ & $2$ &$-{\frac {25}{4}} {\lambda}^{2}-{\frac {15}{4}} d\lambda-\frac{23}2
\lambda-\frac32 d-{\frac {2693}{576}}
$\\$3$ & $3$ &$0
$\\$3$ & $4$ &$-{\frac {55}{8}} {\lambda}^{3}-{\frac {4375}{256}} {\lambda}^{2}-5 
d{\lambda}^{2}-{\frac {168149}{12288}} \lambda-{\frac {1075}{192}} d
\lambda-{\frac {3024601}{884736}}-{\frac {2645}{2304}} d
$\\$4$ & $3$ &${\frac {143}{192}} {\lambda}^{2}+{\frac {5}{16}} d\lambda+{\frac {
11729}{9216}} \lambda+{\frac {4801}{8192}}+{\frac {23}{96}} d
$\\$4$ & $4$ &$0
$\\$5$ & $4$ &$-{\frac {175}{192}} {\lambda}^{3}-{\frac {13427}{6912}} {\lambda}^{2
}-{\frac {125}{192}} d{\lambda}^{2}-{\frac {2125}{3456}} d\lambda-{
\frac {1505639}{1105920}} \lambda-{\frac {24750911}{79626240}}-{
\frac {44165}{331776}} d
$\\\bottomrule
\end{tabular}

\end{table*}
\begin{table*}
 \centering
 \renewcommand{\arraystretch}{1.8}\renewcommand{\baselinestretch}{1.4}
  \caption{Coefficients $w^\text{i}_{ik}(y)$ up to the order $q=4$.}
\begin{tabular}{ll*{1}{>{\normalsize\PBS\raggedright}p{0.8\textwidth}}}
\toprule
$i$ & $k$ & $w^\text{i}_{ik}$\\
\midrule
$0$ & $0$ &$-4 {y}^{2}+4+8 \lambda
$\\$1$ & $0$ &$0
$\\$1$ & $1$ &$ \left( 12 \lambda-4 \right) y-3 {y}^{3}
$\\$2$ & $0$ &$\frac32 \lambda+{\frac {21}{8}}+ \left( 9 \lambda-3 \right) {y}^{2}-{
\frac {15}{8}} {y}^{4}
$\\$2$ & $1$ &$0
$\\$2$ & $2$ &$ \left( 10 \lambda-{\frac {35}{12}} \right) {y}^{2}-\frac74 {y}^{4}
$\\$3$ & $0$ &$0
$\\$3$ & $1$ &$ \left( -\frac38 \lambda+{\frac {55}{16}} \right) y+ \left( {\frac {75}{4
}} \lambda-{\frac {95}{16}} \right) {y}^{3}-{\frac {105}{32}} {y}^{5
}
$\\$3$ & $2$ &$0
$\\$3$ & $3$ &$ \left( {\frac {245}{48}} \lambda-{\frac {3395}{1152}} \right) {y}^{3
}-{\frac {63}{64}} {y}^{5}
$\\$4$ & $0$ &${\frac {25}{16}} {\lambda}^{3}+{\frac {115}{96}} {\lambda}^{2}+{
\frac {1057}{2304}} \lambda+{\frac {115}{1152}}+ \left( -{\frac {9}{
32}} \lambda+{\frac {165}{64}} \right) {y}^{2}+ \left( {\frac {375}{
32}} \lambda-{\frac {475}{128}} \right) {y}^{4}-{\frac {245}{128}} {
y}^{6}
$\\$4$ & $1$ &$0
$\\$4$ & $2$ &$ \left( \frac52 {\lambda}^{2}+{\frac {365}{128}} \lambda+{\frac {30815}{
9216}} \right) {y}^{2}+ \left( {\frac {945}{64}} \lambda-{\frac {2905
}{512}} \right) {y}^{4}-{\frac {315}{128}} {y}^{6}
$\\$4$ & $3$ &$0
$\\$4$ & $4$ &$ \left( {\frac {1645}{576}} \lambda-{\frac {27223}{13824}} \right) {y
}^{4}-{\frac {693}{1280}} {y}^{6}
$\\\bottomrule
\end{tabular}

\end{table*}
\begin{table*}
 \centering
 \renewcommand{\arraystretch}{1.8}\renewcommand{\baselinestretch}{1.4}
  \caption{Coefficients $G_{li}$ up to the order $q=4$.\label{G_table}}
\begin{tabular}{ll*{1}{>{\normalsize\PBS\raggedright}p{0.8\textwidth}}}
\toprule
$l$ & $i$ & $G_{li}$\\
\midrule
$1$ & $0$ &$-16
$\\$1$ & $1$ &$0
$\\$1$ & $2$ &$-12
$\\$1$ & $3$ &$0
$\\$1$ & $4$ &$-{\frac {25}{8}} {\lambda}^{2}-{\frac {895}{48}} \lambda-{\frac {
11305}{1152}}
$\\$2$ & $1$ &$6
$\\$2$ & $2$ &$0
$\\$2$ & $3$ &${\frac {55}{2}} \lambda+{\frac {409}{24}}
$\\$2$ & $4$ &$0
$\\$3$ & $2$ &$-5 \lambda-{\frac {19}{6}}
$\\$3$ & $3$ &$0
$\\$3$ & $4$ &$-5 {\lambda}^{2}-{\frac {995}{64}} \lambda-{\frac {34361}{4608}}
$\\$4$ & $3$ &${\frac {55}{48}} \lambda+{\frac {869}{1152}}
$\\$4$ & $4$ &$0
$\\$5$ & $4$ &$-{\frac {25}{48}} {\lambda}^{2}-{\frac {2195}{3456}} \lambda-{\frac{20819}{103680}}$
\\\bottomrule
\end{tabular}

\end{table*}

\label{lastpage}

\end{document}